%% file: main_all.tex
\RequirePackage[mathlines]{lineno} % Display line numbers
\documentclass[a4paper,11pt]{article}
\pdfoutput=1 % if you are submitting a pdflatex (i.e. if you have
\usepackage{jheppub} % for details on the use of the package, please
\usepackage[T1]{fontenc} % if needed

\usepackage{threeparttable}%This package is used for tablenotes
\usepackage{multirow}
\usepackage{subfigure}
\usepackage{hyperref}
\usepackage{float}

\let\oldequation\equation
\let\oldendequation\endequation
\renewenvironment{equation}
  {\linenomathNonumbers\oldequation}
  {\oldendequation\endlinenomath}
\def \jpsi {J/\psi}

\def \gev  {~\mbox{GeV}}
\def \gevc {~\mbox{GeV/$c$}}
\def \gevcc{~\mbox{GeV/$c^2$}}
\def \mev  {~\mbox{MeV}}

\begin{document}
%\setrunninglinenumbers
%\begin{linenumbers}

\title{\boldmath Search for a massless particle beyond the Standard Model in the $\Xi^0\to\Lambda + \text{invisible}$ decay}
\collaborationImg{\includegraphics[height=30mm,angle=90]{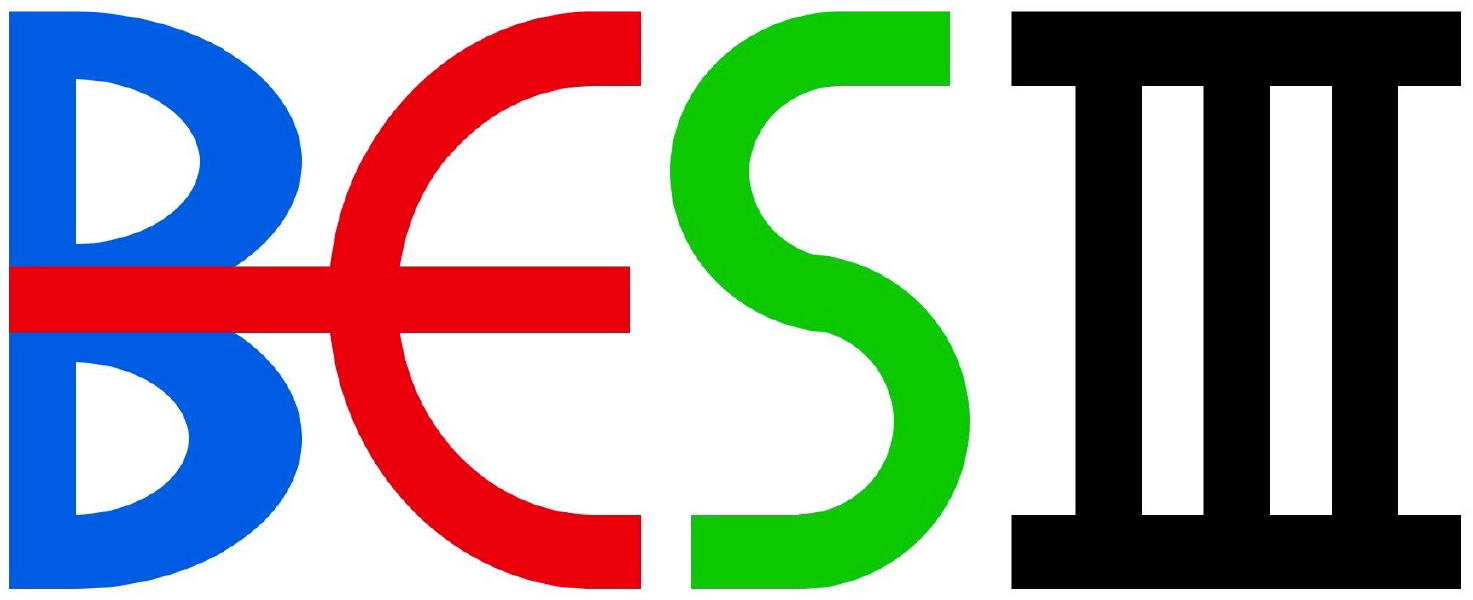}}
\collaboration{The BESIII collaboration}
\emailAdd{besiii-publications@ihep.ac.cn}

\abstract{
A search for a massless beyond-standard-model particle is performed in the decay $\Xi^{0}\to\Lambda+\text{invisible}$ using $(1.0087 \pm 0.0044)\times 10^{10}$ $\jpsi$ events collected with the BESIII detector at the BEPCII collider. No significant signal is observed and the upper limit on the branching fraction $\mathcal{B}(\Xi^{0}\to\Lambda+\text{invisible})$ is set to be $2.3 \times 10^{-4}$ at the $90\%$ confidence level. 
This is the first search for a flavor-changing neutral current process with missing energy in $\Xi^0$ decays. Throughout this paper, charge-conjugate processes are always implied. 
%Throughout this paper, charge-conjugate processes are always implied. This is the first search for a flavor-changing neutral current process with missing energy in $\Xi^0$ decays.
}


\arxivnumber{}
\newcommand{\BESIIIorcid}[1]{\href{https://orcid.org/#1}{\hspace*{0.1em}\raisebox{-0.45ex}{\includegraphics[width=1em]{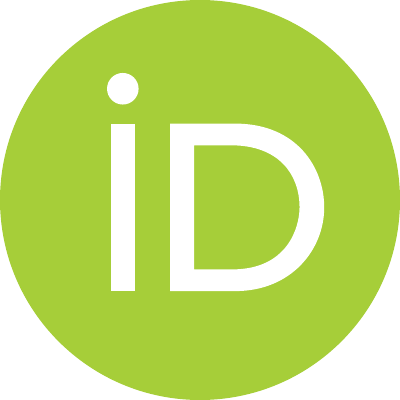}}}}
\maketitle
\flushbottom

%\linenumbers

\section{INTRODUCTION}
\label{sec:introduction}
\begin{figure*}[htbp] \centering
	\setlength{\abovecaptionskip}{-1pt}
	\setlength{\belowcaptionskip}{10pt}

        \subfigure[]
        {\includegraphics[width=0.3\textwidth]{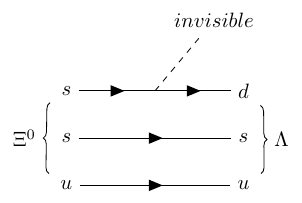}}
        \subfigure[]
        {\includegraphics[width=0.3\textwidth]{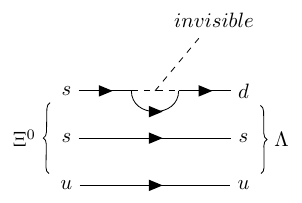}}
        \subfigure[]
        {\includegraphics[width=0.3\textwidth]{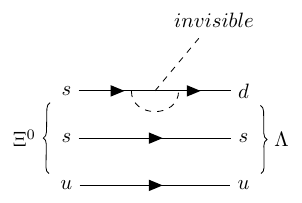}}\\

	\caption{The possible Feynman diagrams for decay $\Xi^{0}\to\Lambda+\text{invisible}$ in NP models: (a) tree-level Feynman diagram (such as QCD axion) and (b)(c) one-loop Feynman diagrams (such as dark photon).}
	\label{fig:fm}
\end{figure*}

\hspace{1.5em}
The flavor-changing neutral-current (FCNC) transitions $s\to d\bar\nu\nu$ are loop-induced and Glashow-Iliopoulos-Maiani~(GIM) suppressed in the Standard Model~(SM), with the expected branching fractions~(BFs) below $10^{-11}$~\cite{Buchalla:1995vs}. If new invisible particles beyond the SM contribute, the BFs for some FCNC hyperon decays could be enhanced to $10^{-4}$~\cite{Tandean:2019tkm}. The search for such decays is therefore a sensitive probe for new physics~(NP). 
%NP particles such as the QCD axion or the dark photon could appear via tree-level or one-loop Feynman diagrams for FCNC decay $\Xi^{0}\to\Lambda+\text{invisible}$, as shown in Fig.~\ref{fig:fm}.
NP particles such as the QCD axion or the dark photon could contribute to the FCNC decay $\Xi^{0}\to\Lambda+\text{invisible}$ via tree-level or one-loop processes, as shown in Fig.~\ref{fig:fm}.

The QCD axion is predicted by the Peccei-Quinn mechanism to resolve the strong CP problem \cite{Peccei:1977hh, Peccei:1977ur, Weinberg:1977ma, Wilczek:1977pj}.
The QCD axion mass scales inversely with its decay constant, $f_a$, which is constrained to be much greater than $10^{6}$\gev~\cite{Georgi:1986df}. Consequently, the mass of the axion is expected to be significantly below the electronvolt scale. In this regime, the axion lifetime exceeds the age of the Universe, so it is a viable dark matter~(DM) candidate~\cite{Preskill:1982cy, Abbott:1982af, Dine:1982ah}. In FCNC transitions relevant for axion searches, hyperon decays such as $\Xi^{0} \to \Lambda + \text{invisible}$ could provide good sensitivity to specific couplings between the axion and quarks~\cite{MartinCamalich:2020dfe}.
The possibility of a long-range gauge force associated with DM is considered, in the form of a new unbroken abelian field, which has been termed the $U(1)_{D}$ dark photon, denoted as $\gamma'$~\cite{Dobrescu:2004wz, Gabrielli:2016cut, He:2017zzr, Ackerman:2008kmp}. If the $U(1)_{D}$ symmetry remains unbroken, its associated gauge boson will not acquire any mass. Although a massless dark photon lacks direct couplings to SM fermions, it can still induce FCNC processes through higher-dimensional operators.

In this paper, we search for the beyond-standard-model massless particle in the two-body decay $\Xi^0\to\Lambda+\text{invisible}$, where the $\Xi^0$ candidate is identified by tagging a $\bar{\Xi}^0$ decaying to $\bar{\Lambda}\pi^{0}$ on the recoiling side. We employ approximately 10 million $\Xi^0\bar{\Xi}^0$ pairs from $(1.0087 \pm 0.0044)\times 10^{10}$ $\jpsi$ events~\cite{bes3:totJpsiNumber} recorded with the BESIII detector. Throughout this paper, charge-conjugate processes are always implied unless explicitly specified.

\section{BESIII DETECTOR AND MONTE CARLO SIMULATION}
\label{sec:detector}
\hspace{1.5em}
The BESIII detector~\cite{bes3:2010detector} records symmetric $e^+e^-$ collisions provided by the BEPCII storage ring~\cite{Yu:IPAC2016-TUYA01} in the center-of-mass energy range from 1.84 to 4.95\gev, with a peak luminosity of $1.1 \times 10^{33}\;\text{cm}^{-2}\text{s}^{-1}$ achieved at $\sqrt{s} = 3.773\;\text{GeV}$. BESIII has collected large data samples in this energy region~\cite{Ablikim:2019hff, EcmsMea, EventFilter, Liao:2025lth}. The cylindrical core of the BESIII detector covers 93\% of the full solid angle and consists of a helium-based multilayer drift chamber~(MDC), a plastic scintillator time-of-flight system~(TOF) and a CsI(Tl) electromagnetic calorimeter~(EMC), which are all enclosed in a superconducting solenoidal magnet providing a 1.0~T magnetic field. The magnetic field was 0.9~T in 2012, corresponding to 11\% of the total $J/\psi$ data. The solenoid is supported by an octagonal flux-return yoke with resistive plate counter muon identification modules interleaved with steel.
%The acceptance of charged particles and photons is 93\% over $4\pi$ solid angle.
The charged-particle momentum resolution at $1\gevc$ is $0.5\%$, and the ${\rm d}E/{\rm d}x$ resolution is $6\%$ for electrons from Bhabha scattering. The EMC measures photon energies with a resolution of $2.5\%$ ($5\%$) at $1\gev$ in the barrel (end-cap) region. The time resolution in the TOF barrel region is 68~ps, while that in the end-cap region is 110~ps. The end-cap TOF system was upgraded in 2015 using multigap resistive plate chamber technology, providing a time resolution of 60~ps, and this upgraded system was utilized for 87\% of the data in this analysis~\cite{etof, etof2, etof3}.

The Monte Carlo (MC) simulated data samples are produced with a {\sc geant4}-based~\cite{geant4} software package, which includes the geometric description of the BESIII detector~\cite{detvis,Li:2024pox,Song:2025pnt,geo1, geo2} and the detector response, are used to determine detection efficiencies and to estimate backgrounds. The simulation models the beam energy spread and initial-state radiation in the $e^+e^-$ annihilations with the generator {\sc kkmc}~\cite{ref:kkmc, ref:kkmc2}. The inclusive MC sample includes both the $\jpsi$ resonance production and the continuum processes simulated in {\sc kkmc}~\cite{ref:kkmc, ref:kkmc2}. All particle decays are modeled with {\sc evtgen}~\cite{ref:evtgen, ref:evtgen2} using BFs either taken from the Particle Data Group~(PDG)~\cite{ParticleDataGroup:2024cfk}, when available, or otherwise estimated with {\sc lundcharm}~\cite{ref:lundcharm, ref:lundcharm2}. Final-state radiation from charged final-state particles is incorporated using the {\sc photos} package~\cite{photos2}.
To obtain the tag efficiency of $\bar{\Xi}^0\to\bar{\Lambda}\pi^0$, the single tag~(ST) MC sample $\jpsi\to\Xi^{0}(\to\Lambda+\text{anything})\bar{\Xi}^{0}(\to\bar{\Lambda}\pi^{0})$ is generated according to its helicity decay amplitudes~\cite{BESIII:2023drj}. The signal MC sample for the decay $\jpsi\to\Xi^{0}(\to\Lambda+\text{invisible})\bar{\Xi}^{0}(\to\bar{\Lambda}\pi^{0})$ is generated according to its helicity decay amplitudes, where the decay-asymmetry parameter of $\Xi^{0}\to\Lambda+\text{invisible}$ is assumed to be the same as that of $\Xi^{0}\to\Lambda\pi^{0}$ decay. The subsequent decays $\Lambda\to p\pi^{-}$, $\bar{\Lambda}\to \bar{p}\pi^{+}$ and $\pi^{0}\to\gamma\gamma$ are generated with the phase space model.

\section{SINGLE TAG EVENT SELECTION}
\label{sec:single}
\hspace{1.5em}
The analysis is performed using the BESIII offline software system~\cite{bes3:boss705}. The signal $\Xi^{0}\to\Lambda+\text{invisible}$ is searched for with $\Lambda\to p\pi^{-}$ from $\Xi^{0}\bar{\Xi}^{0}$ pairs at the center-of-mass energy of $\sqrt{s} = 3.097\gev$. The $\bar{\Xi}^0$ candidates reconstructed through the dominant decay mode $\bar{\Xi}^0 \to \bar{\Lambda} \pi^0$ are denoted as ST side, with $\mathcal{B}_{\bar{\Xi}^0 \to \bar{\Lambda} \pi^0}=(99.524\pm0.012)\%$ from the PDG~\cite{ParticleDataGroup:2024cfk}. The recoiling system opposite to the ST side in the space is defined as the double tagged~(DT) side and treated as the signal. The BF of the signal decay can be calculated as:
\begin{eqnarray}
\mathcal{B}_{\Xi^{0}\to\Lambda+\text{invisible}}=\frac{N_{\text{DT}}\times\epsilon_{\text{ST}}}{N_{\text{ST}}\times\epsilon_{\text{DT}}\times\mathcal{B}_{\Lambda\to p\pi^{-}}},
\label{eq:BF}
\end{eqnarray}
where $N_{\text{ST}}~(N_{\text{DT}})$ is the observed ST~(DT) yield, $\epsilon_{\text{ST}}~(\epsilon_{\text{DT}})$ the corresponding detection efficiency, and $\mathcal{B}_{\Lambda\to p\pi^{-}}=(64.1\pm0.5)\%$ the BF of $\Lambda\to p\pi^{-}$ from the PDG~\cite{ParticleDataGroup:2024cfk}.

Events are required to have zero net charge and at least two charged tracks. All charged tracks must be detected in the MDC within the range of $|\text{cos}\theta|<0.93$, where $\theta$ is the polar angle with respect to the axis of symmetry of the MDC. The distance of closest approach from the interaction point must be less than 20 cm along the beam direction.

Particle identification (PID) is applied by combining the measurements of d$E$/d$x$ in the MDC and the time-of-flight in the TOF to calculate the likelihoods $\mathcal{L}(h)$ for each hadron hypothesis $h=p, K, \pi$.
The $\pi$ candidates are required to satisfy $\mathcal{L}(\pi)>0$, $\mathcal{L}(\pi)>\mathcal{L}(K)$, while the $p$ candidates must satisfy $\mathcal{L}(p)>0$, $\mathcal{L}(p)>\mathcal{L}(K)$ and $\mathcal{L}(p)>\mathcal{L}(\pi)$. Due to the low momenta of pions~(less than $0.4\gevc$) and protons~(less than $0.9\gevc$), the identification efficiencies for pions and protons exceed 99\%, with misidentification probabilities less than 1\%. At least one $\pi$ candidate and one $p$ candidate are required for each event.

Photon candidates are selected from showers deposited in the EMC with energies of $E_{\gamma} > 25\mev$ in the barrel ($|\cos\theta|<0.80$) and $E_{\gamma} > 50\mev$ in the end-cap ($0.86 < |\text{cos}\theta| < 0.92$).  Showers must be separated from the extrapolated positions of any charged track by at least $10^{\circ}$ ($20^{\circ}$ for the anti-proton track) and occur within 700~ns after the event start time. At least two photon candidates are required per event. The $\pi^{0}$ candidates are reconstructed by looping over all pairs of good photons and their invariant mass is required to lie in the range of (0.115, 0.150)$\gevcc$, which is approximately a $\pm2\sigma$ window around the nominal mass of $\pi^{0}$. The $\chi^2_{\gamma\gamma}$ from a one-constraint~(1C) kinematic fit constraining their invariant mass to the known $\pi^{0}$ mass~\cite{ParticleDataGroup:2024cfk} must be less than 25. The updated $\pi^0$ four-momentum from the 1C kinematic fit is used in later calculations.

The $\bar{\Lambda}$ candidates are reconstructed via the $\bar{p}\pi^{+}$ final state. Primary and secondary vertex fits are performed to all $\bar{p}\pi^{+}$ candidates, where the primary vertex is the production vertex of the $\bar{\Lambda}$ and the secondary vertex is its decay vertex. The $\chi^{2}$ variables of primary and secondary vertex fits are both required to be less than 100. The decay length $L$ divided by its uncertainty $\sigma_{L}$ obtained from the secondary vertex fit must be greater than 2. The $\bar{p}\pi^{+}$ invariant mass $M_{\bar{p}\pi^{+}}$, obtained after the secondary vertex fit, is required to lie within the interval $(1.111, 1.120)\gevcc$, which is about a $\pm3\sigma$ window around the nominal mass of $\Lambda$. The $\bar{\Lambda}$ candidates satisfying all the above requirements are designated as good $\bar{\Lambda}$ candidates. At least one good $\bar{\Lambda}$ candidate is required. The updated $\bar{\Lambda}$ four-momentum, obtained by the secondary vertex fit, is used in all subsequent steps.

\begin{figure*}[htbp]
\centering
\setlength{\abovecaptionskip}{-1pt}
	\setlength{\belowcaptionskip}{10pt}
	\includegraphics[width=7.0cm]{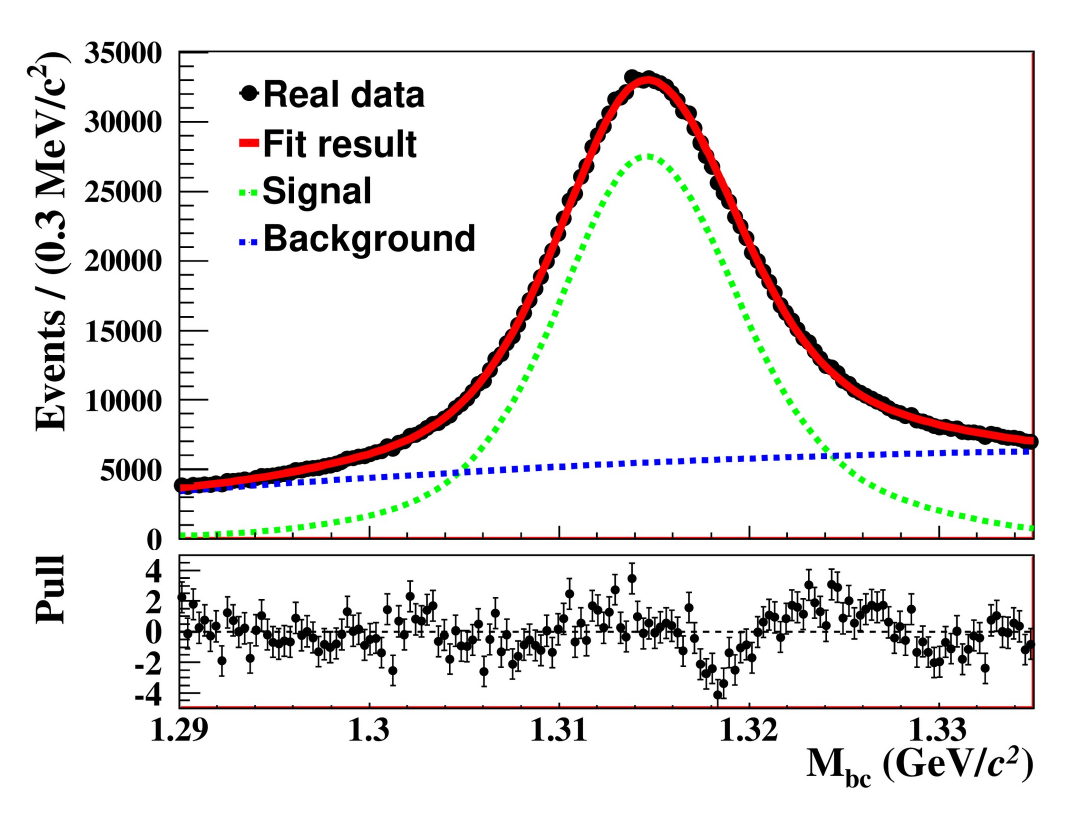}
\caption{The fit on the $M_{\rm{bc}}$ from the ST side. The black dots with error bars represent the data and the red line represents the  total fit, whose signal component is shown with the green dashed line and the background component with the blue dashed line. The bottom panel shows the pull values, defined as $Pull^{i}=\frac{N_{\text{data}}^{i}-N_{\text{fit}}^{i}}{\sigma_{\text{data}}^{i}}$, where $N_{\text{data}}^{i}$ is the number of data events in the $i$-th bin, $N_{\text{fit}}^{i}$ is the corresponding total number of fitted events, and $\sigma_{\text{data}}^{i}$ is the statistical uncertainty of the data.}
\label{fig:STfit}
\end{figure*}

The $\bar{\Xi}^{0}$ is reconstructed with the combination of $\bar{\Lambda}$ and $\pi^{0}$ candidates using their updated four-momentum after the fits. The best ST $\bar{\Xi}^{0}$ candidate is determined with minimum $\Delta E=|E_{\bar{\Lambda}\pi^{0}}-E_{\rm{cms}}/2|$ value from all possible $\bar{\Lambda}$ and $\pi^{0}$ combinations, where $E_{\bar{\Lambda}\pi^{0}}$ is the total energy of the $\bar{\Lambda}$ and $\pi^{0}$ candidates, and $E_{\rm{cms}}$ is the center-of-mass energy. $|M_{\bar{\Lambda}\pi^{0}}-M_{\bar{\Xi}^{0}_{\rm{PDG}}}|<0.02\gevcc$ is required for $\bar{\Xi}^{0}$ candidate, where $M_{\bar{\Lambda}\pi^{0}}$ is the invariant mass of $\bar{\Lambda}$ and $\pi^{0}$, and $M_{\bar{\Xi}^{0}_{\rm{PDG}}}$ is the nominal mass of $\bar{\Xi}^{0}$ from the PDG~\cite{ParticleDataGroup:2024cfk}. The ST yield is extracted from a binned maximum likelihood fit to the distribution of $M_{\rm{bc}}$, defined as
\begin{eqnarray}
    M_{\rm{bc}}=\sqrt{E^{2}_{\rm{cms}}/4c^{4}-|\vec{P}|_{\bar{\Lambda}\pi^{0}}^{2}/c^{2}},
\label{eq:Mbc}
\end{eqnarray}
where $\vec{P}_{\bar{\Lambda}\pi^{0}}$ is the total momentum of the $\bar{\Lambda}$ and $\pi^{0}$ candidates. The signal process $\jpsi\to\bar{\Xi}^{0}(\to\Lambda\pi^{0})\Xi^{0}(\to X)$ is dominant, while the background distribution is smooth and well described by a polynomial. The signal shape is described by the MC-simulated shape convolved with a Gaussian function, which represents the resolution difference between the data and MC simulation. A second-order Chebyshev polynomial describes the background shape in the fit. We obtain $(1.254\pm0.004)\times 10^{6}$ ST events in the signal region, as shown in Fig.~\ref{fig:STfit}, where the uncertainty is statistical only.

%%%%%%%%%%%%%%%%%%%%%%%%%%%%%%%%%%%%%
%%%%%%%%%%%%%%%%%%%%%%%%%%%%%%%%%%%%%
The ST detection efficiencies are evaluated using the signal MC sample, and are found to be $(8.15 \pm 0.01)\%$. As a consistency check, the BF of $\jpsi\to\bar{\Xi}^{0}\Xi^{0}$ is measured to be $(1.212\pm0.004)\times10^{-3}$ according to the observed ST yield and the corresponding ST efficiency, which is consistent with the previous BESIII measurements~\cite{BESIII:2023drj, BESIII:2016nix}.

\section{DOUBLE TAG EVENT SELECTION}
\label{sec:analysis}
\hspace{1.5em}
After a $\bar{\Xi}^{0}$ candidate is identified on the ST side, the signal decay $\Xi^{0}\to\Lambda+\text{invisible}$ is probed in the DT side, where the final state consists of $p, \pi^{-}$ and an invisible particle. Exactly two additional charged particles are required, and they must be identified as a proton and a pion, which are used to reconstruct the $\Lambda$ candidate. The selection requirements for protons and pions on the DT side are the same as those for the ST side, as introduced in Section~\ref{sec:single}.

Two two-constraint~(2C) kinematic fits are performed under the hypothesis of $\jpsi\to p\bar{p}\pi^{+}\pi^{-}\pi^{0}+\rm{invisible}$, which impose four-momentum conservation and assign the $\pi^{0}$ mass to the photon pair. In the first 2C kinematic fit, the invisible particle is assumed to be massless, while in the second fit, it is assigned the nominal $\pi^{0}$ mass. The $\chi^{2}_{\text{2C}}$, obtained from the first 2C kinematic fit, must be less than 8.5. The four-momenta of the invisible particle, $p$ and $\pi^{-}$ candidates on the DT side are obtained by this fit.  To suppress the background from $\Xi^{0}\to\Lambda\pi^{0}$, the $\chi^{2}_{\text{2C}\pi^{0}}$ of the second 2C kinematic fit must be greater than $\chi^{2}_{\text{2C}}$. The invariant mass of $p\pi^-$ and the invisible particle, denoted as $M_{p\pi^{-}+\text{inv.}}$, is required to be in the range of $(1.310, 1.322)\gevcc$. For the background processes $\Xi^{0}\to\Lambda\gamma$ and $\Xi^{0}\to\Lambda\pi^{0}$, the missing photon carries the energy and momentum. Typically, the missing photon is either outside the detector~($|\text{cos}\theta|>0.92$) or located in the gaps between the barrel and the two end caps ($0.80<|\text{cos}\theta|<0.86$). Therefore, the polar angle of the invisible particle must satisfy $|\text{cos}\theta_\text{inv.}|<0.8$. To suppress the background $\jpsi\to\Sigma^{0}(\to\Lambda\gamma)\bar{\Sigma}^{0}(\bar{\Lambda}\gamma)$, the missing momentum $|\vec{P}_\text{inv.}|$ from the first 2C kinematic fit, is required to be $|\vec{P}_\text{inv.}| > 0.15\gevc$.

If the number of photon candidates exceeds two, a five-constraint~(5C) kinematic fit is performed under the hypothesis of $\jpsi\to p\bar{p}\pi^{+}\pi^{-}\pi^{0}\gamma$, where the $\gamma$ is one of the photons not originating from the $\pi^{0}$ on the ST side. We loop over all such photons and require the minimum $\chi_{3\gamma}^{2}$ to be greater than 1000. In addition, if the number of photon candidates exceeds three, a six-constraint~(6C) kinematic fit is performed under the hypothesis of $\jpsi\to p\bar{p}\pi^{+}\pi^{-}\pi^{0}\gamma\gamma$, where the $\gamma\gamma$ are two photons not originating from the $\pi^{0}$ on the ST side. We loop over all such photons and require the minimum $\chi_{4\gamma}^{2}$ to be greater than 1000.

The selection criteria on $M_{\rm{bc}}$, $\chi_{\text{2C}}^{2}$ and $M_{p\pi^{-}+\text{inv.}}$ are optimized by maximizing the $\text{Punzi significance}=\frac{\epsilon}{1.5+\sqrt{B}}$~\cite{Punzi:2003bu}, where $\epsilon$ denotes the signal efficiency obtained from signal MC sample and $B$ is the number of background events obtained from background MC samples.

\section{SIGNAL EXTRACTION}
\label{sec:DTyield}
\begin{figure*}[htbp]
\centering
\setlength{\abovecaptionskip}{-1pt}
\setlength{\belowcaptionskip}{10pt}
\includegraphics[width=7.0cm]{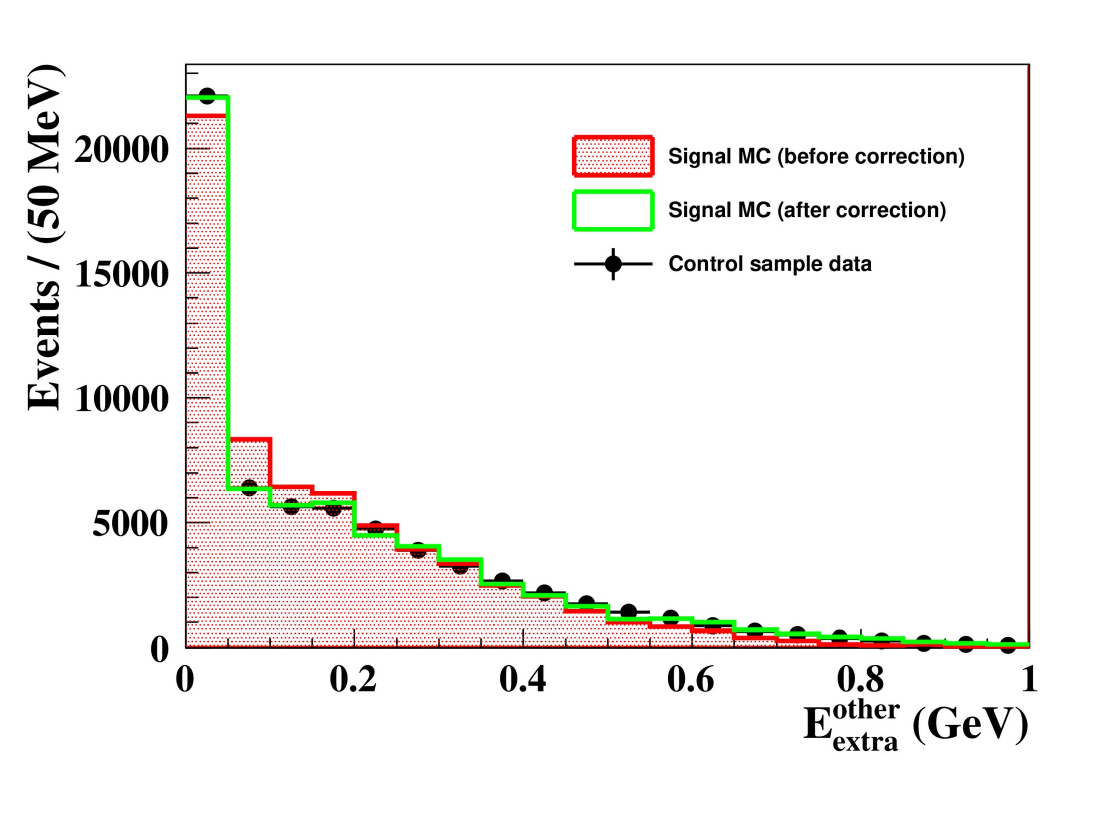}
\caption{The $E_{\text{extra}}^{\text{other}}$ distribution. The red histogram represents the Signal MC before the correction, the green histogram represents the Signal MC after the correction, and the black dots with error bars represent the $\jpsi\to\bar{\Xi}^{0}(\to\bar{\Lambda}\pi^{0})\Xi^{0}(\to\Lambda\pi^{0})$ control sample.}
\label{fig:eother}
\end{figure*}

\hspace{1.5em}
After the full event selection, the residual background is dominated by $J/\psi\to\bar{\Xi}^{0}(\to\bar{\Lambda}\pi^{0})\,\Xi^{0}(\to\Lambda\pi^{0})$ and $J/\psi\to\bar{\Sigma}^{*0}(\to\bar{\Lambda}\pi^{0})\,\Sigma^{*0}(\to\Lambda\pi^{0})$. Since the invisible particle should not deposit energy in the EMC, the total energy of all EMC showers excluding the ST $\pi^{0}$, denoted as $E_{\text{extra}}$, provides powerful discrimination. The $E_{\text{extra}}$ can be divided into two parts
\begin{eqnarray}
E_{\text{extra}}= E_{\text{extra}}^{\text{DT}\pi^0}+E_{\text{extra}}^{\text{other}},
\label{eq:eextra}
\end{eqnarray}
where $E_{\text{extra}}^{\text{DT}\pi^0}$ is the energy of the $\pi^{0}$ on the DT side, expected to be zero for signal events. The value of $E_{\text{extra}}^{\text{DT}\pi^0}$ in the background is obtained from MC simulation, whose modelling of the photon and electron interactions with the detector materials is sufficiently accurate. The $E_{\text{extra}}^{\text{other}}$ comes from other contributions, including from unrelated sources such as detector and electronic noises. After suppression of the induced showers through the isolation-angle requirements, the interactions of the $\bar{p}$ track with the detector materials are estimated to dominate the residual energy. Owing to the limitation of the GEANT4 in modelling the anti-proton interactions with detector materials, the raw simulated $E_{\text{extra}}^{\text{other}}$ disagrees with the distribution from the data, as shown in Fig.~\ref{fig:eother}. Therefore, the shape of $E_{\text{extra}}^{\text{other}}$ is corrected using a data-driven approach based on a $\jpsi\to\bar{\Xi}^{0}(\to\bar{\Lambda}\pi^{0})\Xi^{0}(\to\Lambda\pi^{0})$
control sample. The contribution of $E_{\text{extra}}^{\text{other}}$ is assigned by sampling a random value from a shape template extracted from the control data sample, binned in the momentum and the polar angle of the anti-proton.

%%%%%%%%%%%%%%%%%%%%%%%%%%%%%%%%%%%%%
%%%%%%%%%%%%%%%%%%%%%%%%%%%%%%%%%%%%%
%%%%%%%%%%%%%%%%%%%%%%%%%%%%%%%%%%%%%
\begin{figure*}[htbp]
\centering
\setlength{\abovecaptionskip}{-1pt}
\setlength{\belowcaptionskip}{10pt}
\includegraphics[width=7.0cm]{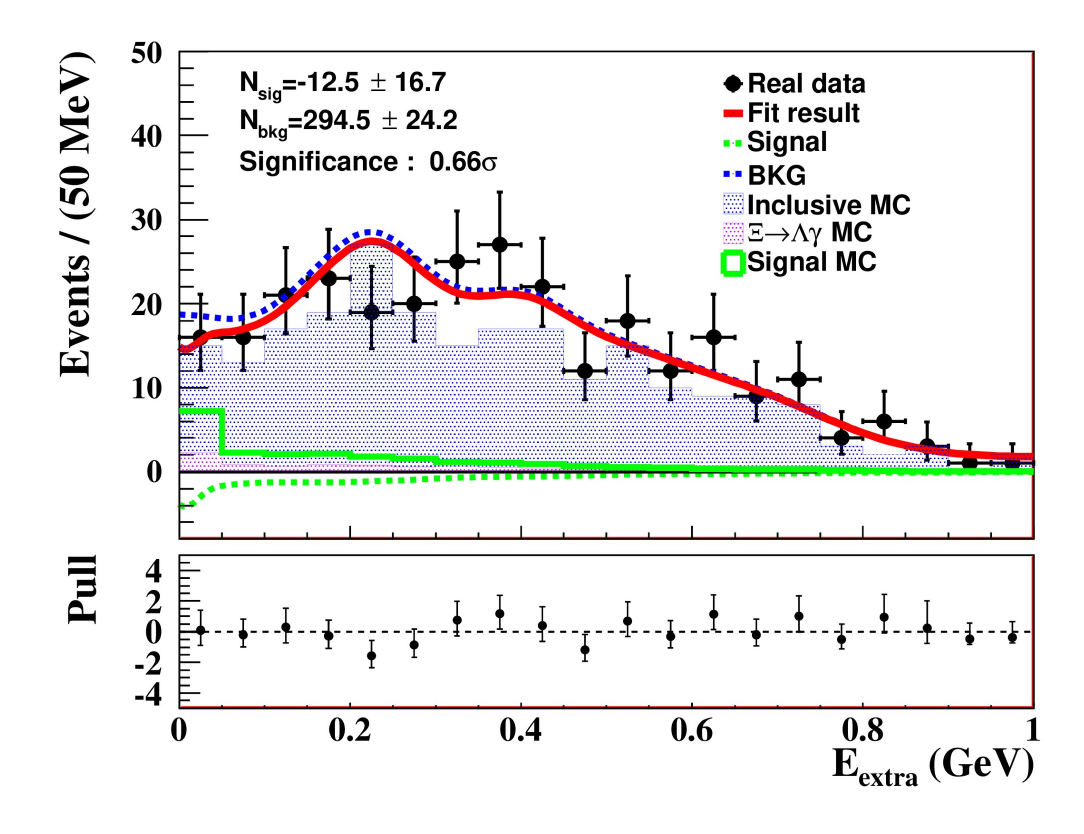}
\caption{Fit on the $E_{\text{extra}}$ distribution. The black dots with error bars represent the data sample, the red line represents the fitted curve, the green dashed line represents the signal component, and the blue dashed line represents the background component. The green histogram represents the signal MC sample~(scaled according to the obtained upper limit of its BF), the blue histogram represents the inclusive MC sample, and the violet histogram represents the $\Xi^0\to\Lambda\gamma$ MC sample. The bottom panel shows the pull values defined as $Pull^{i}=\frac{N_{\text{data}}^{i}-N_{\text{fit}}^{i}}{\sigma_{\text{data}}^{i}}$, where $N_{\text{data}}^{i}$ is the number of data events in the $i$-th bin, $N_{\text{fit}}^{i}$ is the corresponding total number of fitted events, and $\sigma_{\text{data}}^{i}$ is the statistical uncertainty of the data.}
\label{fig:DTfit}
\end{figure*}
%%%%%%%%%%%%%%%%%%%%%%%%%%%%%%%%%%%%%
A binned extended maximum likelihood fit is performed to the corrected distribution of $E_{\text{extra}}$ to determine the signal yields $N_{\text{fit}}$, as shown in Fig.~\ref{fig:DTfit}. The probability density functions of signal and background are derived from the shape of the signal MC and inclusive MC simulation after the data-driven correction. The fitted yield $N_{\text{fit}}$ is $-12.5 \pm 16.7$. The peaking background from $\Xi^{0}\to\Lambda\gamma$, with the expected yield $N_{\text{Peaking}}=8.5$ from a $\Xi^{0}\to\Lambda\gamma$ MC sample, mimics the signal but is absent from the inclusive MC sample. The DT signal yield $N_{\text{DT}}$ is obtained by subtracting the peaking background $N_{\text{Peaking}}$ from the fitted yield $N_{\text{fit}}$. The result indicates no significant excess of observed signal above the expected backgrounds.

\section{SYSTEMATIC UNCERTAINTIES}
\label{sec:systematic}
\hspace{1.5em}
The DT analysis technique cancels most systematic uncertainties related to the ST selection. The remaining systematic uncertainties are divided into multiplicative uncertainties and additive uncertainties.

The multiplicative systematic uncertainties arise from the MC generator, the intermediate BF, the peaking background, and the selection requirements on $\chi^{2}_{\text{2C}}$, $\cos\theta_{\text{inv.}}$, $|\vec{P}_\text{inv.}|$, $M_{p\pi^{-}+\text{inv.}}$, $\chi^{2}_{3\gamma}$, and $\chi^{2}_{4\gamma}$. Adding the individual components in quadrature gives a total multiplicative systematic uncertainty of 9.5\%, as summarized in Table~\ref{tab:syst}.
\begin{itemize}
\item \emph{MC generator.} To study the uncertainty arising from the signal MC generator, the decay parameters $\alpha_{\Xi^{0}}$ used in the mDIY model are varied from 0 to -0.5 and +0.5~\cite{BESIII:2023drj}. The largest efficiency difference of 5.0\% between the alternative and nominal signal MC samples is assigned as the uncertainty.
\item \emph{Intermediate BF.} The uncertainty of $\mathcal{B}(\Lambda\to p\pi^{-})$ from the PDG~\cite{ParticleDataGroup:2024cfk} is 0.8\%.
\item \emph{Peaking background.} The uncertainty on the peaking background $\Xi^0\to\Lambda\gamma$ is estimated to be 6.9\%, combining the BF uncertainty quoted from the PDG~\cite{ParticleDataGroup:2024cfk}  with the limited MC statistics.
\item \emph{Requirements on } $\chi^{2}_{\text{2C}}$, $\text{cos}\theta_\text{inv.}$, $|\vec{P}_\text{inv.}|$, $M_{p\pi^{-}+\text{inv.}}$, $\chi^{2}_{3\gamma}$, $\chi^{2}_{4\gamma}$. A control sample $\jpsi\to\bar{\Xi}^{0}\Xi^{0}$, where $\Xi^{0}\to\Lambda\pi^{0}$ and $\bar{\Xi}^{0}\to\Lambda\pi^{0}$, is used to study the selection requirements uncertainties. The same ST selections as those in Section~\ref{sec:single} are applied to the control sample. To suppress the background, further selections $M_{\rm{bc}}\in(1.304, 1.321)\gevcc$ and $\chi_{4\gamma}^{2}<100$ are applied. Specifically, the $\chi_{4\gamma}^{2}<100$ requirement is omitted when studying the uncertainty associated with this specific criterion. Under the same selections, the efficiency differences between the MC and the data are taken as the corresponding uncertainties. The uncertainties from the $\chi^{2}_{\text{2C}}$, $\text{cos}\theta_\text{inv.}$, $|\vec{P}_\text{inv.}|$, $M_{p\pi^{-}+\text{inv.}}$, $\chi^{2}_{3\gamma}$, $\chi^{2}_{4\gamma}$ requirements are found to be $2.1\%$, $1.1\%$, $1.7\%$, $1.4\%$, $1.9\%$ and $1.8\%$, respectively.
\end{itemize}
\begin{table*}[tpb]
\setlength{\abovecaptionskip}{0.0cm}
\caption{Summary of the multiplicative systematic uncertainties for the BF measurements. The total value is calculated by summing up all sources in quadrature.}
  \begin{center}
  \footnotesize
  \newcommand{\tabincell}[2]{\begin{tabular}{@{}#1@{}}#2\end{tabular}}
  \begin{threeparttable}
  \begin{tabular}{l|c c c c c c c}
      \hline \hline
			Sources &  Relative systematic uncertainties~(\%)\\
			\hline
            MC generator & $5.0$\\
            Intermediate BF & $0.8$\\
            Peaking background & $6.9$\\
		      $\chi^{2}_\text{2\rm{C}}$ requirement & $2.1$\\
            $\text{cos}\theta_\text{inv.}$ requirement & $1.1$\\
            $|\vec{P}_\text{inv.}|$ requirement& $1.7$\\
            $M_{p\pi^{-}+\text{inv.}}$ requirement & $1.4$\\
            $\chi^{2}_{3\gamma}$ requirement& $1.9$\\
            $\chi^{2}_{4\gamma}$ requirement& $1.8$\\
    			\hline
                         Total & 9.5\\% & \\
\hline \hline
\end{tabular}
\end{threeparttable}
\end{center}
\label{tab:syst}
\end{table*}

The additive systematic uncertainties mainly come from the bin width, the background shape and the signal shape. Since we perform the binned maximum-likelihood fit on the $E_{\text{extra}}$ distribution using $20\mev$ bin width, the uncertainty arising from the choice of bin width is considered by using alternative bin widths of $10\mev$ and $40\mev$. To consider the influence of the fluctuations caused by the background shape, we change the smoothing parameter ($\rho$ value) of the RooKeysPDF function from 1 to 2 or 3. The uncertainty associated with the signal shape is considered by replacing the nominal model with alternative models in which the decay-asymmetry parameter of $\Xi^{0}\to \Lambda + \rm{invisible}$ is varied between -0.5 and 0.5. We obtain an upper limit~(UL) for each alternative case and take the most conservative UL as the final result.

\section{RESULT}
\label{sec:result}
\hspace{1.5em}
Since no significant signal is observed, we set the UL on $\mathcal{B}(\Xi^{0}\to\Lambda+\text{invisible})$ at the $90\%$ confidence level~(C.L.) after considering the systematic uncertainties. For each fixed signal yield, a BF value is obtained according to Eq.~\ref{eq:BF}, and a likelihood value is obtained from the fitting. We scan the signal yields 200 times by varying the number of events with a step size of 0.3 to obtain the likelihood distribution $\mathcal{L}_{i}$ and the maximum likelihood $\mathcal{L}_{\rm{max}}$. The relative likelihood $\mathcal{L}$ is defined as
\begin{eqnarray}
\mathcal{L}=\frac{\mathcal{L}_{i}}{\mathcal{L}_{\rm{max}}}.
\label{eq:L}
\end{eqnarray}
A Gaussian fit is performed to the distribution of the relative likelihood.
Then we obtain a mean value and an uncertainty of the BF, denoted $\hat{\mathcal{B}}$ and $\sigma_{\mathcal{B}}$. The related likelihood function is
\begin{eqnarray}
\mathcal{L}(\mathcal{B})_{\rm{fit}}\propto \exp[-\frac{(\mathcal{B}-\hat{\mathcal{B}})^{2}}{2\sigma^{2}_{\mathcal{B}}}].
\label{eq:A}
\end{eqnarray}
By following a method~\cite{Liu:2015uha} which incorporates the systematic uncertainties into the UL of the BF, we obtain  the smeared likelihood function,
\begin{eqnarray}
\mathcal{L}(\mathcal{B})_{\rm{smear}}\propto \int_{0}^{1} \exp[-\frac{(\epsilon\mathcal{B}/\hat{\epsilon}-\hat{\mathcal{B}})^{2}}{2\sigma^{2}_{\mathcal{B}}}]\times\frac{1}{\sqrt{2\pi}\sigma_{\rm{sys}}\hat{\epsilon}} \exp[-\frac{(\epsilon-\hat{\epsilon})^{2}}{2\sigma^{2}_{\rm{sys}}\hat{\epsilon}^{2}}]d\epsilon,
\label{eq:B}
\end{eqnarray}
where $\hat{\epsilon}$ is the nominal efficiency and $\sigma_{\rm{sys}}$ denotes the multiplicative systematic uncertainty.

\begin{figure}[htbp] \centering
	\setlength{\abovecaptionskip}{-1pt}
	\setlength{\belowcaptionskip}{10pt}
	\includegraphics[width=7.0cm]{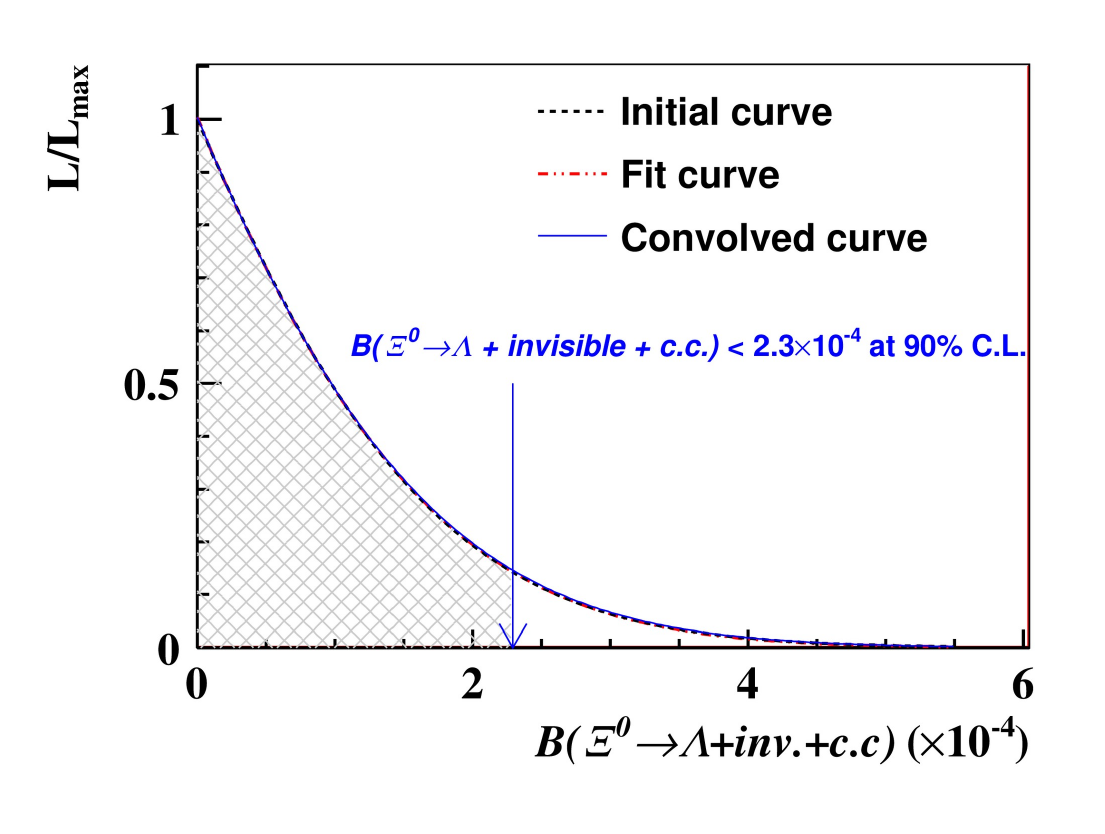}
	\caption{The distributions of the likelihood curves. The dotted-dashed black line is the initial curve, the dotted-dashed red line is the fit curve with a Gaussian function, and the solid blue line is the fit curve convolved with a Gaussian for systematic uncertainties. The blue arrow indicates the UL on BF at the 90\% C.L.}
	\label{fig:limit}
\end{figure}

\begin{figure}[htbp] \centering
	\setlength{\abovecaptionskip}{-1pt}
	\setlength{\belowcaptionskip}{10pt}
	\includegraphics[width=7.0cm]{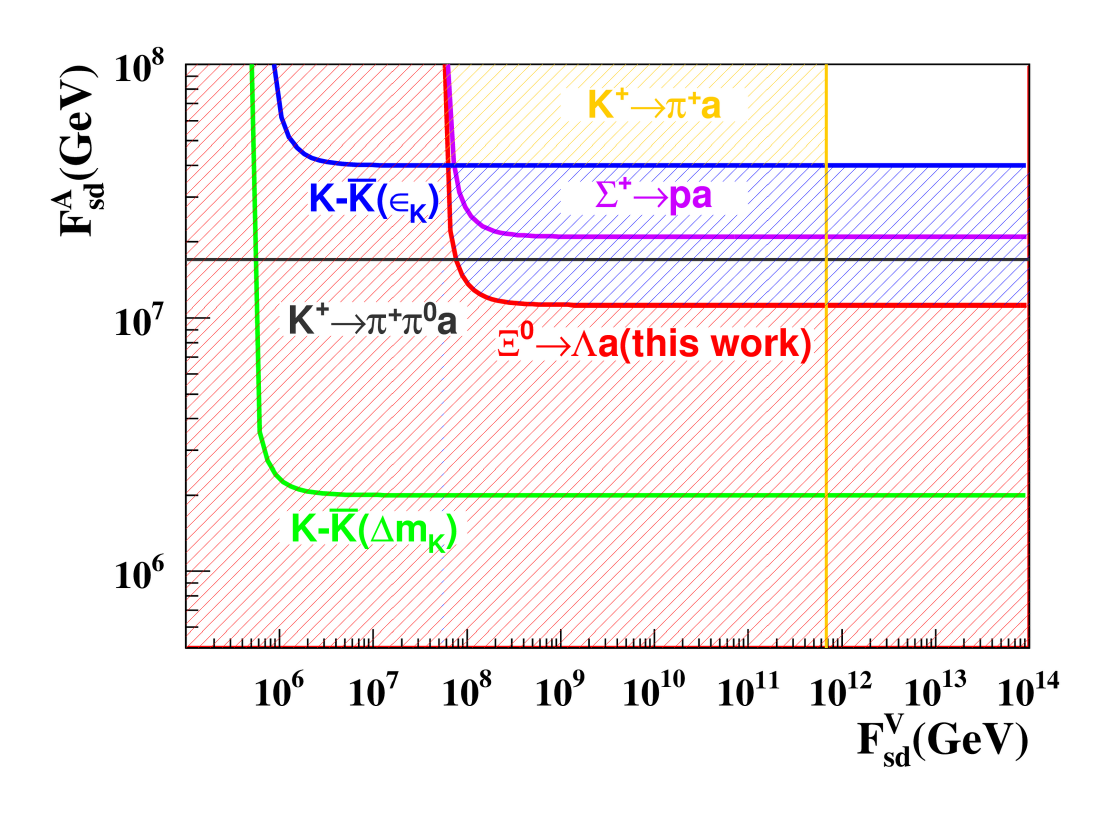}
	\caption{The $90\%$ C.L. exclusion limits of $s\to d$ flavor-violating axion couplings obtained by various experiments, $\Sigma^{+}\to pa$~\cite{BESIII:2023utd}, $K^{+}\to\pi^{+}a$~\cite{E949:2007xyy}, $K^{+}\to\pi^{+}\pi^{0}a$~\cite{E787:2000iwe}, $K-\bar{K}(\Delta m_{K})$~\cite{ParticleDataGroup:2024cfk}, $K-\bar{K}(\epsilon_{K})$~\cite{UTfit:2022hsi}.}
	\label{fig:axion}
\end{figure}

Also considering the additive systematic uncertainties detailed in Section~\ref{sec:systematic}, we obtain the most conservative UL at the $90\%$ C.L. on the BF to be $\mathcal{B}(\Xi^{0}\to\Lambda+\text{invisible})<2.3\times10^{-4}$, as shown in Fig.~\ref{fig:limit}.
%%%%%%%%%%%%%%%%%%%%%%%%
%%%%%%%%%%%%%%%%%%%%%%%%%
Under the hypothesis of a massless dark photon, the maximum allowed BF of $\Xi^{0}\to\Lambda\gamma'$ is $1.2\times10^{-4}$~\cite{Su:2019ipw}, which remains consistent with our experimental bound. For the QCD axion, the vector coupling $F_{sd}^{V}$ is already tightly constrained by $K^{+}\to\pi^{+}a$~\cite{E949:2007xyy} searches, as shown in Fig.~\ref{fig:axion}. However, a lower bound of the axial-vectorial part $F_{sd}^{A}$ is set using the upper limit obtained in this study, which is significantly better than the constraint from $K-\bar{K}$ mixing~($\Delta m_{K}$)~\cite{ParticleDataGroup:2024cfk} and competitive with that from searches for $\Sigma^{+}\to pa$~\cite{BESIII:2023utd}, $K^{+}\to\pi^{+}\pi^{0}a$~\cite{E787:2000iwe} and measurements of the CP-violating parameter $\epsilon_{K}$~\cite{UTfit:2022hsi} in the kaon system, as shown in Fig.~\ref{fig:axion}.
%%%%%%%%%%%%%%%%%%%%%%%%
%%%%%%%%%%%%%%%%%%%%%%%%%
\section{SUMMARY}
\label{sec:summary}
\hspace{1.5em}
A massless particle beyond the SM is searched for in the $\Xi^{0}$ FCNC decay $\Xi^{0}\to\Lambda+\text{invisible}$ for the first time, using $(1.0087 \pm 0.0044)\times 10^{10}$  $\jpsi$ events collected with the BESIII detector. No significant signals are observed above the expected backgrounds.
The UL on BF is set to be $\mathcal{B}(\Xi^{0}\to\Lambda+\text{invisible})<2.3\times10^{-4}$ at the $90\%$ C.L. after taking into account the systematic uncertainties. Throughout this paper, charge-conjugate processes are always implied. 
%This result imposes a stringent limit on the NP models with a massless particle beyond the SM.
This result provides the first experimental constraint on this decay mode and imposes stringent limits on various new physics models involving a massless BSM particle.
\acknowledgments
\hspace{1.5em}\begin{sloppypar}
The BESIII Collaboration thanks the staff of BEPCII (https://cstr.cn/31109.02.BEPC) and the IHEP computing center for their strong support. This work is supported in part by National Key R\&D Program of China under Contracts Nos. 2023YFA1606000, 2023YFA1606704; National Natural Science Foundation of China (NSFC) under Contracts Nos. 11635010, 11935015, 11935016, 11935018, 12025502, 12035009, 12035013, 12061131003, 12192260, 12192261, 12192262, 12192263, 12192264, 12192265, 12221005, 12225509, 12235017, 12342502, 12361141819; the Chinese Academy of Sciences (CAS) Large-Scale Scientific Facility Program; the Strategic Priority Research Program of Chinese Academy of Sciences under Contract No. XDA0480600; CAS under Contract No. YSBR-101; 100 Talents Program of CAS; The Institute of Nuclear and Particle Physics (INPAC) and Shanghai Key Laboratory for Particle Physics and Cosmology; ERC under Contract No. 758462; German Research Foundation DFG under Contract No. FOR5327; Istituto Nazionale di Fisica Nucleare, Italy; Knut and Alice Wallenberg Foundation under Contracts Nos. 2021.0174, 2021.0299, 2023.0315; Ministry of Development of Turkey under Contract No. DPT2006K-120470; National Research Foundation of Korea under Contract No. NRF-2022R1A2C1092335; National Science and Technology fund of Mongolia; Polish National Science Centre under Contract No. 2024/53/B/ST2/00975; STFC (United Kingdom); Swedish Research Council under Contract No. 2019.04595; U. S. Department of Energy under Contract No. DE-FG02-05ER41374
\end{sloppypar}

\newpage
\input{authorlist_2025-12-07.tex}
%\end{linenumbers}
\end{document}

%% file: authorlist_2025-12-07.tex
%% Saved at => 2025-12-07
\begin{sloppypar}M.~Ablikim$^{1}$\BESIIIorcid{0000-0002-3935-619X},
M.~N.~Achasov$^{4,d}$\BESIIIorcid{0000-0002-9400-8622},
P.~Adlarson$^{83}$\BESIIIorcid{0000-0001-6280-3851},
X.~C.~Ai$^{89}$\BESIIIorcid{0000-0003-3856-2415},
C.~S.~Akondi$^{31A,31B}$\BESIIIorcid{0000-0001-6303-5217},
R.~Aliberti$^{39}$\BESIIIorcid{0000-0003-3500-4012},
A.~Amoroso$^{82A,82C}$\BESIIIorcid{0000-0002-3095-8610},
Q.~An$^{79,65,\dagger}$,
Y.~H.~An$^{89}$\BESIIIorcid{0009-0008-3419-0849},
Y.~Bai$^{63}$\BESIIIorcid{0000-0001-6593-5665}
% ...
\\
\vspace{0.2cm}
(BESIII Collaboration)\\
\vspace{0.2cm} {\it
$^{1}$ Institute of High Energy Physics, Beijing 100049, People's Republic of China\\
% ...
$^{4}$ Budker Institute of Nuclear Physics SB RAS (BINP), Novosibirsk 630090, Russia\\
% ...
$^{31}$ INFN Sezione di Ferrara, (A)INFN Sezione di Ferrara, I-44122, Ferrara, Italy; (B)University of Ferrara, I-44122, Ferrara, Italy\\
% ...
$^{39}$ Johannes Gutenberg University of Mainz, Johann-Joachim-Becher-Weg 45, D-55099 Mainz, Germany\\
% ...
$^{63}$ Southeast University, Nanjing 211100, People's Republic of China\\
% ...
$^{65}$ State Key Laboratory of Particle Detection and Electronics, Beijing 100049, Hefei 230026, People's Republic of China\\
% ...
$^{79}$ University of Science and Technology of China, Hefei 230026, People's Republic of China\\
% ...
$^{82}$ University of Turin and INFN, (A)University of Turin, I-10125, Turin, Italy; (B)University of Eastern Piedmont, I-15121, Alessandria, Italy; (C)INFN, I-10125, Turin, Italy\\
$^{83}$ Uppsala University, Box 516, SE-75120 Uppsala, Sweden\\
% ...
$^{89}$ Zhengzhou University, Zhengzhou 450001, People's Republic of China\\

\vspace{0.2cm}
$^{\dagger}$ Deceased\\
% ...
$^{d}$ Also at the Novosibirsk State University, Novosibirsk, 630090, Russia\\
% ...
}
\end{sloppypar}
%% ends here %%